\documentclass[conference]{IEEEtran}

\usepackage[utf8]{inputenc}

\usepackage{amsmath,amssymb,amsfonts}
\usepackage{graphicx}
\usepackage{textcomp}
\usepackage{subcaption}
\usepackage{gensymb}
\usepackage{float}
\usepackage{algorithm,algcompatible,algpseudocode}
\usepackage{array}
\usepackage{diagbox}
\usepackage[table]{xcolor}
\usepackage{multirow}
\usepackage{multicol}
\usepackage{tabularx, makecell}
\usepackage{tabularray}
\usepackage{ctable}

\def\BibTeX{{\rm B\kern-.05em{\sc i\kern-.025em b}\kern-.08em
    T\kern-.1667em\lower.7ex\hbox{E}\kern-.125emX}}
\begin{document}

\title{Deep Feature Learning for Wireless Spectrum Data}

\author{\IEEEauthorblockN{Ljupcho Milosheski, Gregor Cerar, Bla\v{z} Bertalani\v{c}, Carolina Fortuna and Mihael Mohor\v{c}i\v{c}}\\
\textit{Jozef Stefan Institute, Ljubljana, Slovenia}\\
\{ljupcho.milosheski, gregor.cerar, blaz.bertalanic, carolina.fortuna, miha.mohorcic\}@ijs.si}

\maketitle

\begin{abstract}

In recent years, the traditional feature engineering process for training machine learning models is being automated by the feature extraction layers integrated in deep learning architectures. In wireless networks, many studies were conducted in automatic learning of feature representations for domain-related challenges.
However, most of the existing works assume some supervision along the learning process by using labels to optimize the model.
In this paper, we investigate an approach to learning feature representations for wireless transmission clustering in a completely unsupervised manner, i.e. requiring no labels in the process. We propose a model based on convolutional neural networks that automatically learns a reduced dimensionality representation of the input data with 99.3\% less  components compared to a baseline principal component analysis (PCA). We show that the automatic representation learning is able to extract fine-grained clusters containing the shapes of the wireless transmission bursts, while the baseline enables only general separability of the data based on the background noise.

\end{abstract}


\section{Introduction}

The introduction of machine learning (ML) algorithms in wireless communication has led to improvement of the existing and development of completely new solutions when sufficient data is available. Some examples include modulation classification~\cite{o2018over, rajendran2018deep}, radio technology classification~\cite{fontaine2019towards, fonseca2021radio}, anomaly detection~\cite{feng2017anomaly}, and device fingerprinting~\cite{merchant2018deep, robinson2020dilated}. ML techniques that rely on manually engineered features from the data are gradually being replaced by deep learning (DL) algorithms which are able to extract more relevant features as an integral part of their training process \cite{riyaz2018deep}. Features extracted using deep-learning models appear to contain more meaningful information \cite{fontaine2019towards} and allow scaling to larger datasets while at the same time improving the accuracy \cite{robinson2020dilated}.

Although these DL models provide unmatched accuracy in domain-based classification tasks, they require large amounts of labeled data for training, i.e. larger than classical machine learning algorithms. Such large amounts of training data for instance on radio spectrum usage are typically collected and made available for the research community from real-world environment either using wireless testbed networks such as LOG-a-TEC \cite{vsolc2015low} or crowd-sourcing initiative such as ElectroSense \cite{rajendran2017electrosense}. However, labeling radio spectrum data requires domain specialists with good knowledge of the operating environment and understanding of wireless technologies, making it an expensive and erroneous process. To address this issue, usage of unsupervised/semi-supervised \cite{feng2017anomaly}, \cite{o2017semi} models is emerging as an alternative, but still under-explored approach.

In this paper, we adapt and propose an architecture for learning feature representations of wireless transmissions from spectrograms in a completely unsupervised manner. In the absence of similar approach for direct comparison, we use the principal components analysis (PCA) as a baseline automatic representation learning approach. The proposed architecture was originally developed for feature learning from color images, known as DeepCluster \cite{caron2018deep}. We adapt this architecture to the domain environment and prove it is a worthy alternative for training a model that outperforms the baseline in the extraction of features that describe and distinguish the spectrogram patterns of different wireless transmission technologies. Considering the lower amount of content dynamics of the spectrograms compared to the color images, we propose a  methodology for selecting the number of dimensions that contain the relevant features in the representation provided by convolutional neural networks (CNNs).

The main contributions of this work are as follows:
\begin{itemize} 
    \item We propose a CNN-based model that automatically learns a reduced dimensionality representation of the input data with 99.3\% less components compared to baseline PCA.
    \item We prove that the proposed CNN-based representation learning is able to extract features that are representing actual transmissions, while PCA can learn only general representations that characterize the background noise.
    \item We develop a methodology for evaluating the quality of the provided features with regards to their clustering tendency, complementary to the clustering quality assessment. The introduction of such evaluation offers additional insight for the selection of the number of clusters and number of dimensions of the reduced feature space, the two critical parameters of the proposed architecture.
\end{itemize}

The rest of the paper is structured as follows. Section \ref{sec:related} analyzes the related work. Section  \ref{sec:Architecture} elaborates on the feature representation learning using DL while Section \ref{sec:methodology} elaborates on the experimental methodology, including the  feature development and evaluation metrics. Section \ref{sec:Results} presents and discusses the experimental results. Finally, Section \ref{sec:conclusions} concludes the paper.

\section{Related work}
\label{sec:related}

Regarding the state of the art feature representation learning approaches available in wireless communications, we identified two related lines of work: supervised feature learning and feature learning incorporating unsupervised architectures. The later can be completely unsupervised or semi-supervised. 

\subsection{Supervised feature learning}

Considering the amount of research works, supervised DL architectures have well established usage for the domain-related problems.
For device fingerprinting, as one of the main tasks in the domain, high classification accuracy is achieved (above 92\%) in \cite{merchant2018deep}, \cite{robinson2020dilated}. The ability of the CNN to encode relevant features is also proven in modulation classification tasks \cite{o2018over}, \cite{rajendran2018deep} where various types of CNN-based architectures are used. Supervised solutions for wireless technology classification achieving high accuracy are proposed in \cite{fontaine2019towards}, \cite{fonseca2021radio}.
It is clear that when classification problem is being addressed and big amount of labeled data is available, CNN-based solutions achieve top performance. But, providing big labeled spectrum data, as was discussed before, is an expensive and erroneous task. 

\subsection{Unsupervised and semi-supervised feature learning}
General unavailability of labelled spectrum data is constraining the usage of the supervised approaches. Thus, efforts are invested in resolving this problem by using architectures that require only small section of the data to be labeled, compromising the accuracy.

In \cite{kuzdeba2021transfer}, dilated causal convolutional (DCC) architecture is used in an unsupervised auto-encoder configuration to learn features from an unlabeled dataset. Small part of the data is labeled and used for tuning the last layers of the network in supervised configuration. They show that the auto-encoder successfully learns the general features of the data. 

In \cite{o2017semi}, an auto-encoder is compared to semi-supervised bootstrapping of sparse representation for modulation classification problem. Authors visually show that using the semi-supervised approach provides better features compared to the unsupervised approach and generalizes better on unseen data, but no quantitative support is provided.

There are also usages with completely unsupervised implementation.
In \cite{feng2017anomaly}, automatic feature learning is proposed with an auto-encoder network for the task of anomaly detection in spectrum data. The network is compared with linear and robust PCA and is shown to better extract features and provides better accuracy of anomaly detection. But this is still a marginal case because the task is a binary classification.

In our work, we aim towards completely automatic representation learning from large amount of unlabeled radio spectrum data for the purpose of clustering, when multiple types of spectrum activities are existing.

\begin{figure*}[!t]
    \centering
    \includegraphics[width=\textwidth]{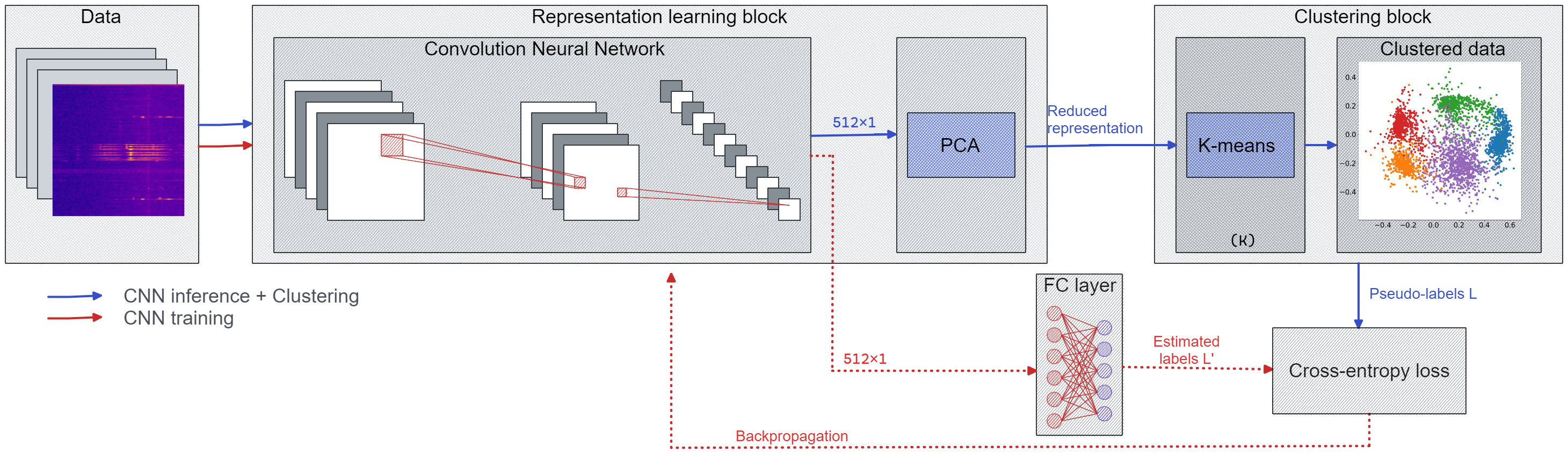}
    \caption[width=\textwidth]{Architecture of the automatic feature learning system.}
    \label{fig:FigureArchitectureDiagram}
\end{figure*}

\section{Feature Representation Learning} 
\label{sec:Architecture}

We propose a CNN-based feature learning and clustering architecture as depicted in Figure~\ref{fig:FigureArchitectureDiagram}. It was inspired by and adapted from the existing DeepCluster model, originally proposed in \cite{caron2018deep} for RGB image features learning. The architecture design contains a representation learning block and a clustering block. The representation learning contains a CNN block followed by PCA performing automatic learning of reduced dimensionality feature representation. The clustering block then processes the data provided by the representation learning block. It is relevant to consider that the same dimensionality reduction could be achieved by using one additional fully connected layer (FCN) after the CNN. However, we use PCA because it allows automatic feature ranking based on the explained variance ratio (EVR) as an integral part of the PCA. The ranking is made in the same Cartesian space in which the K-means is working, allowing for better explainability of the developed models.

Compared to the original DeepCluster model \cite{caron2018deep}, we made the following adaptations:
\begin{itemize}
	\item Rather than using VGG \cite{simonyan2014very} as a DL architecture, we selected ResNet18 \cite{he2016deep} motivated by the performance improvement in a use case involving spectrum data in \cite{o2018over}. Thus, we achieve similar performance while reducing the complexity of the models.
	\item We customized ResNet input and output layers according to the shape of the images and the number of classes. In our case, the input spectrogram images have only one channel, while the ResNet was originally designed for 3-channel RGB images.
\end{itemize}

During the training process of the automatic feature representation learning using the architecture depicted in Figure~\ref{fig:FigureArchitectureDiagram}, a feedback loop is used as shown with a dotted line. It consists of a fully connected classification layer attached to CNN. This layer generates estimated labels used as a ground truth. During the iterative training process they are compared to the pseudo-labels generated by the K-means clustering and the difference is propagated back to guide the training. More specifically,  clustering and CNN weights training are performed in an alternating manner. In the initial phase, the K-means clustering on the output of randomly initialized CNN provides initial cluster assignments which are used as temporary pseudo-labels (L) for the first epoch of training of the CNN. The improved CNN is then used for features extraction in the next iteration, that subsequently with the new clustering provide new temporal labels. The clustering-training sequence completes one training iteration. The procedure stops when the predefined number of iterations (training epochs) is reached. 

It is important to note that through its iterative training the proposed representation learning approach includes a tight coupling between the values of the CNN weights, the size $N$ of the PCA components and the number of clusters $K$ as summarized in the first line of Table~\ref{tab:architectures}. Using this architecture, all these dimensions need to be optimized simultaneously and they influence each other through the feedback loop. However, in the final application, only the optimized system consisting of the representation learning is needed. Two possible ways of utilizing the developed representation learning model are:
\begin{enumerate}
    \item Feature extractor for clustering which provides ability to discover new devices, by varying the parameters of the used clustering algorithm.
    \item Transmission classification for already discovered number of classes using fully connected layer at the output of the CNN.
\end{enumerate}
\begin{table}
\centering
\caption{Adaptable parameters for the Baseline and CNN-based learning architectures.}
\centering
    \begin{tabular}{|l|c|c|}\hline
    \diagbox[width=10em]{Architecture}{Function \\ block}&
    Representation learning & Clustering \\ \hline
    CNN-based & \multicolumn{2}{c}{CNN + PCA (n=1..N) (k=1..K)} \vline \\
    \hline
    Baseline & PCA (n=1..N) & k (K=2..30) \\
    \hline
\end{tabular}

\label{tab:architectures}
\end{table}

The equivalent baseline architecture not involving the CNN learning component and the dotted training loop in Figure~\ref{fig:FigureArchitectureDiagram} can be optimized in a sequential manner: fist optimizing the PCA as a representation learning method and then optimizing the K-means as clustering method. The equivalent optimization parameters are summarized in the second line of Table \ref{tab:architectures}. The baseline system employs a flattening block that  reorders the elements of the input matrix into a single row and feeds them to PCA to learn a representation.

\section{Methodology}
\label{sec:methodology}

\subsection{Training and Evaluation Data} \label{sec:Data}

The dataset used for the performance analysis consists of 15 days of radio spectrum measurements acquired in the LOG-a-TEC testbed at a sampling rate of 5 power spectral density measurements per second using 1024 FFT bins in the 868\,MHz license-free (shared spectrum) band with a 192\,kHz bandwidth. Details of the acquisition process and a subset of data can be found in \cite{vsolc2015low}. The acquired data has a matrix form of $1024 \times M$, where $M$ is the number of measurements over time. 

\begin{figure}[htbp]
\centering
\noindent{\includegraphics[width=\columnwidth]{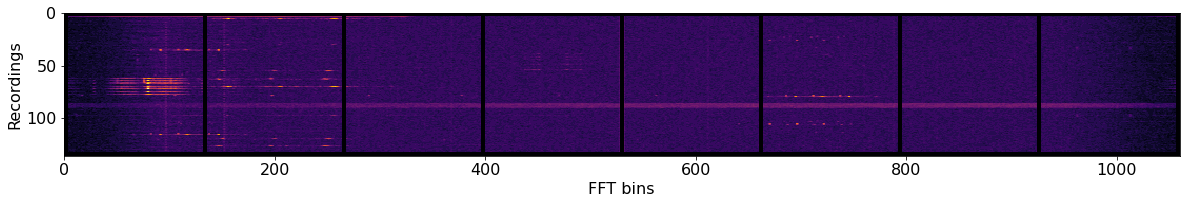}}
\caption[width=\columnwidth]{Sample of 8 spectrogram segments from the data.}
\label{fig:SampleData}
\end{figure}

The complete data-matrix was segmented into non-overlapping square images (spectrograms) along time and frequency (FFT bins) for a window size $W=128$. An example of such segmentation containing 8 square images is shown in Figure~\ref{fig:SampleData}, corresponding to the image resolution of 25.6 seconds (128 measurements taken at 5 measurements per second) by 24 kHz. The window size is large enough to contain any single type of activity and small enough to avoid having too many activities in a single image while also having in mind computational cost. Dividing the entire dataset of 15 days using $W=128$ and zero overlapping, produces $423,904$ images of $128 \times 128$ pixels. Additionally, the pixel values are scaled to [0,1].

\subsection{Optimization of the Representation Learning}
\label{sec:parameters}
CNN-based and baseline approaches can be optimized along two dimensions: the number of PCA components that should be used in the representation and the number of clusters for the K-means. For the baseline model, the two parameters are independent, meaning that the representation learning function is not affected by the number of clusters that will be later used on the obtained feature vectors. On the other side, for the CNN-based architecture, changes in the number of clusters affect the representation learning. This is because the number of clusters should always be the same as the number of classes at the output of the CNN during the learning process, so it inflicts changes on the representation learning block. This means that varying the number of clusters should also be considered when choosing the number of dimensions for the representation learning with the CNN-based model.

It is unfeasible to study the influence of individual CNN weights on the learnt representation and cluster quality due to their large numbers. They are optimized in a black-box manner during the training process consisting of 200 training epochs. This number was determined empirically by observing the convergence of the loss function.

\subsection{Evaluation}
\label{sec:evaluation}
As an evaluation metric for choosing the dimensionality of the representation for both models we use EVR \cite{Jolliffe1986pca}. EVR is a measure of how much of the variation in the feature space is assigned to each of the principal components after performing PCA.

We analyze the quality of the representation for clustering purposes employing visual assessment of tendency (VAT) \cite{bezdek2002vat}. This method produces matrix visualisation of the dissimilarity of randomly selected subset of samples based on their pairwise euclidean distances. The samples are ordered in such a way that groups that are closely located in the feature space, according to the distance metric, appear as dark squares along the diagonal of the matrix. Implementation wise, we used an improved version of VAT (i.e., iVAT) which provides better visualization than the standard one.

We also evaluate the quality of the clustering, performed on the extracted features, by using the Silhouette score metric. In this way we provide quantification of how well the clusters are distinguished for the analysed models.

Using these metrics, we evaluate and explore the representation learning capabilities of both approaches and their applicability for clustering. First we analyze a histogram of samples for the formed clusters based on the frequency sub-band resulting from image segmentation that each sample comes from. These plots provide information on whether the learned feature representation used for the clustering is correlated to the location of the samples along the frequency axis. Then we plot the average of the samples assigned to a single cluster. This provides an insight into the actual spectrogram content that is specific for the formed clusters.

\section{Experimental results} \label{sec:Results}

\subsection{Learning with the PCA baseline approach}
\label{sec:LearningBaseline}
\begin{figure}[htbp]
    \begin{subfigure}[b]{\columnwidth}
        \centering
        \includegraphics[width=\columnwidth]{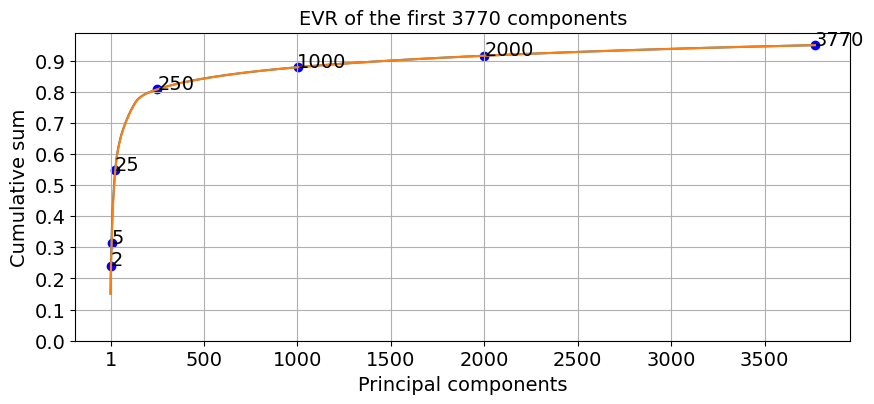}
        \caption{Cumulative sum of EVR for the baseline representation.}
        \label{fig:baselineCumSum}
    \end{subfigure}
    \begin{subfigure}[b]{0.24\columnwidth}
        \centering
        \includegraphics[width=\columnwidth]{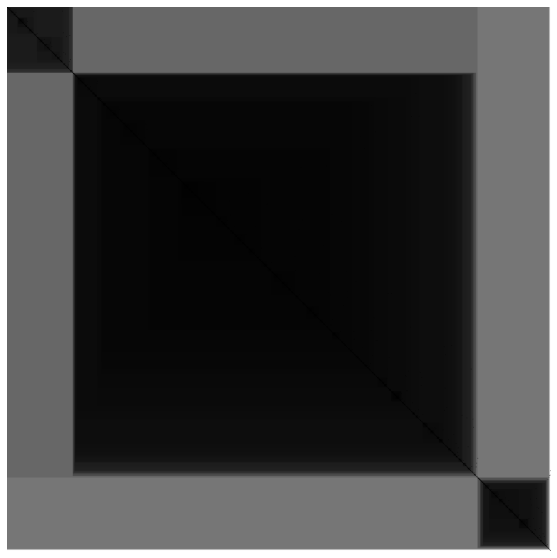}
        \caption{VAT-2}
        \label{fig:baselineVAT2}
    \end{subfigure}
    \begin{subfigure}[b]{0.24\columnwidth}
        \centering
        \includegraphics[width=\columnwidth]{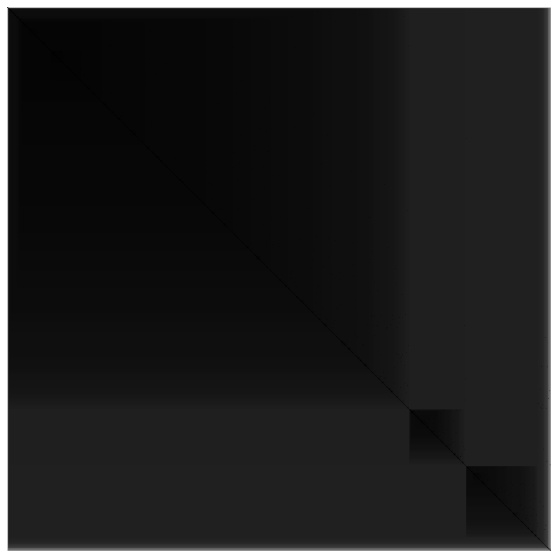}
        \caption{VAT-5}
        \label{fig:baselineVAT5}
    \end{subfigure}
    \begin{subfigure}[b]{0.24\columnwidth}
        \centering
        \includegraphics[width=\columnwidth]{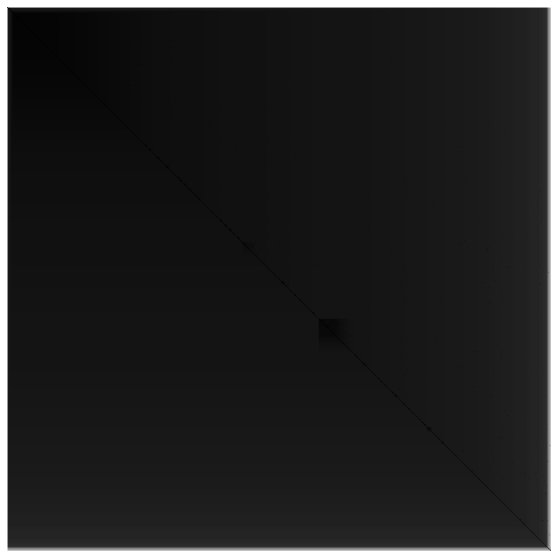}
        \caption{VAT-25}
        \label{fig:baselineVAT25}
    \end{subfigure}
    \begin{subfigure}[b]{0.24\columnwidth}
        \centering
        \includegraphics[width=\columnwidth]{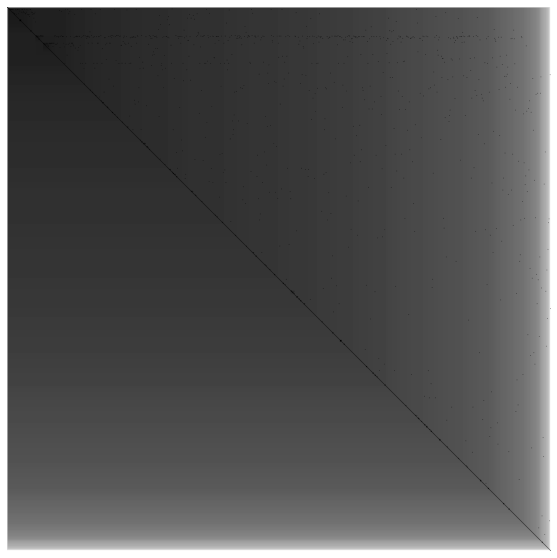}
        \caption{VAT-250}
        \label{fig:baselineVAT250}
    \end{subfigure}
    \begin{subfigure}[b]{0.24\columnwidth}
        \centering
        \includegraphics[width=\columnwidth]{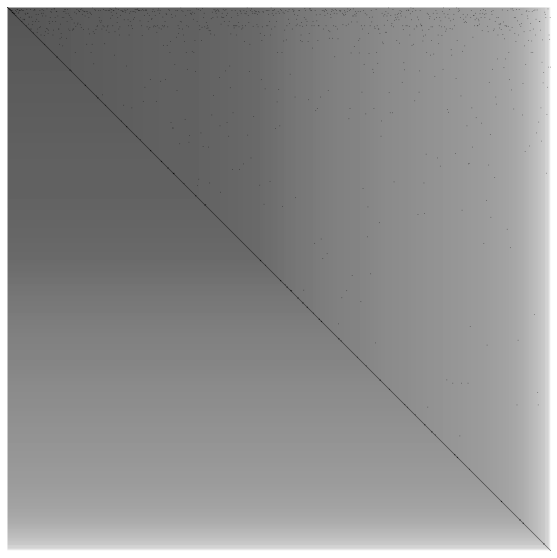}
        \caption{VAT-1000}
        \label{fig:baselineVAT1000}
    \end{subfigure}
    \begin{subfigure}[b]{0.24\columnwidth}
        \centering
        \includegraphics[width=\columnwidth]{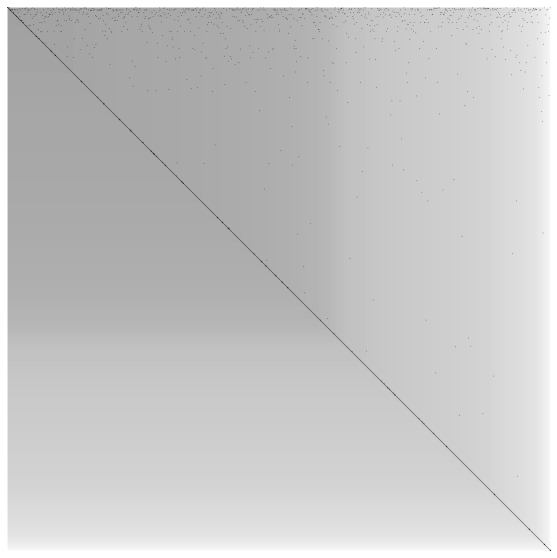}
        \caption{VAT-2000}
        \label{fig:baselineVAT2000}
    \end{subfigure}
    \begin{subfigure}[b]{0.24\columnwidth}
        \centering
        \includegraphics[width=\columnwidth]{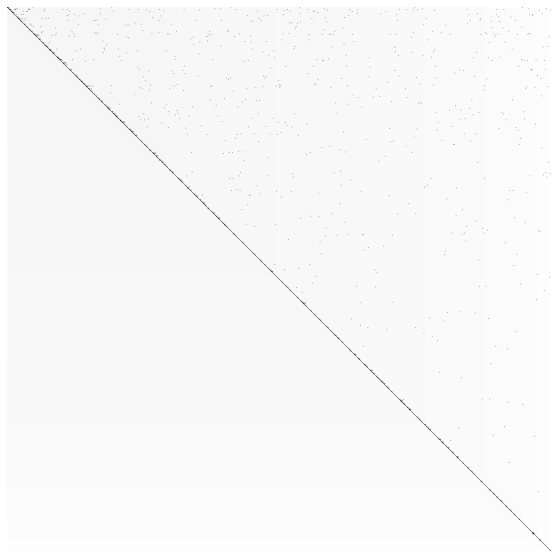}
        \caption{VAT-3770}
        \label{fig:baselineVAT3770}
    \end{subfigure}
\caption{Evaluation of the baseline representation learning.}
\label{fig:RepresentationLearningBaseline}
\end{figure}

In Figure \ref{fig:RepresentationLearningBaseline}, we present the evaluation of the learnt representation according to EVR and VAT metrics discussed in Section \ref{sec:evaluation}. Figure~\ref{fig:baselineCumSum} shows EVR of the features learnt by the baseline representation learning block consisting of PCA only, followed by the VAT plots in Figures~\ref{fig:RepresentationLearningBaseline}b--h. The plots correspond to 7 different PCA-based representation learning models, configured for different number of components selected in a way to evaluate the feature vectors with wide range of different dimensions.

It can be seen that by keeping 95\% of the variance ratio in the PCA learned representation provides feature vectors of dimension 1x3770, which is around 23\% of the flattened single sample input of 1x16384. Although the learned representation has reduced dimensions by more than four times compared to the flattened input, we still have a high dimensionality representation. The VAT plots in Figure~\ref{fig:RepresentationLearningBaseline}b--h show that the baseline approach learns representation with weak clustering tendency for all cases, except the one when using only the first two components of the feature space. The VAT-2 plot (Figure \ref{fig:baselineVAT2}) of the 2-dimensional feature representation shows the existence of three well separated clusters.

\subsection{Learning with the CNN-based approach}
\label{sec:LearningDL}

\begin{figure}[htbp]
    \begin{subfigure}[b]{\columnwidth}
        \centering
        \includegraphics[width=\columnwidth]{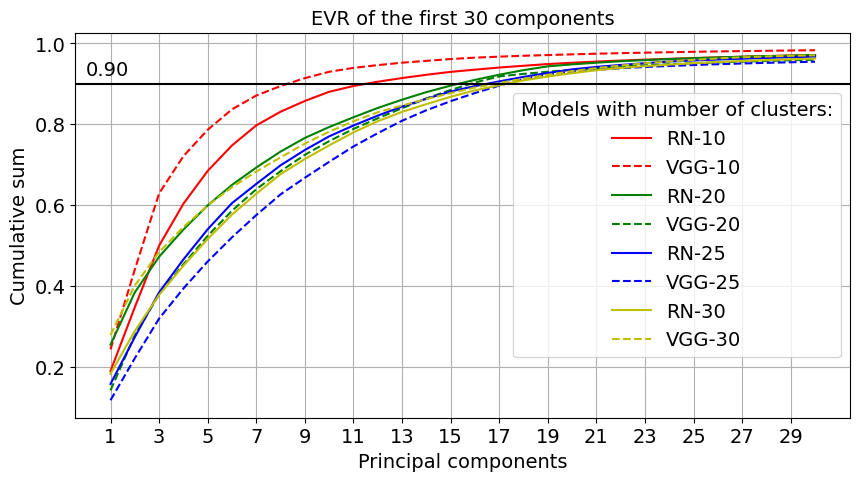}
        \caption{Cumulative sum of EVR for DL-based representations.}
        \label{fig:DLbasedCumSum}
    \end{subfigure}
    \begin{subfigure}[b]{0.24\columnwidth}
        \centering
        \includegraphics[width=\columnwidth]{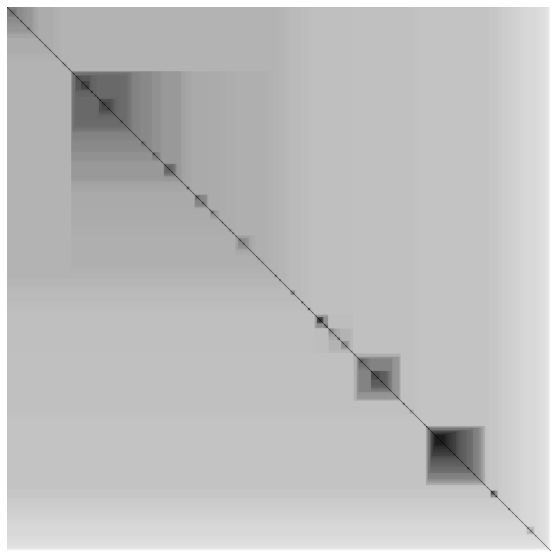}
        \caption{RN-10}
        \label{fig:RN-VAT10}
    \end{subfigure}
    \begin{subfigure}[b]{0.24\columnwidth}
        \centering
        \includegraphics[width=\columnwidth]{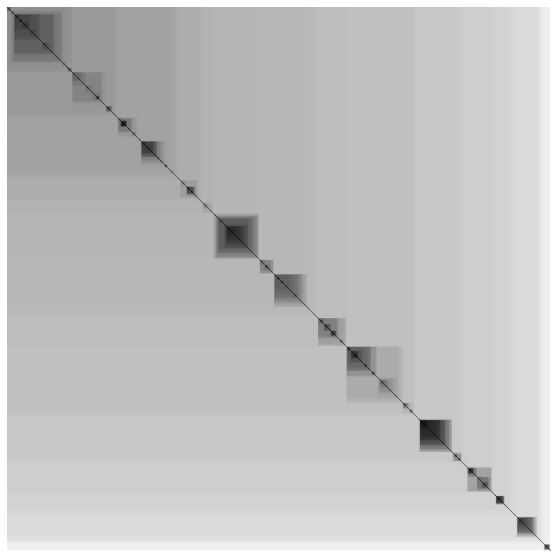}
        \caption{RN-20}
        \label{fig:RN-VAT20}
    \end{subfigure}
    \begin{subfigure}[b]{0.24\columnwidth}
        \centering
        \includegraphics[width=\columnwidth]{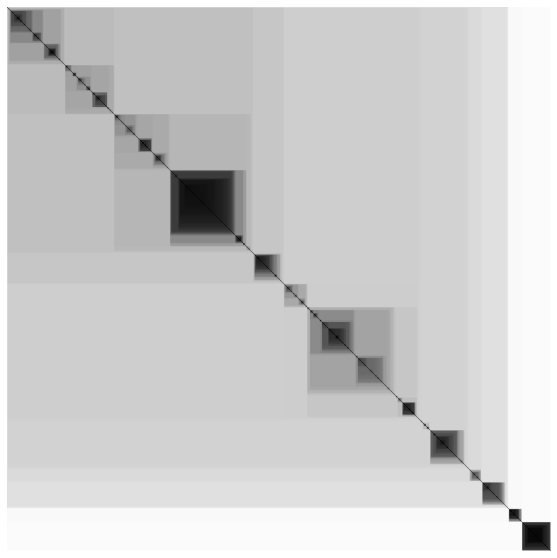}
        \caption{RN-25}
        \label{fig:RN-VAT25}
    \end{subfigure}
    \begin{subfigure}[b]{0.24\columnwidth}
        \centering
        \includegraphics[width=\columnwidth]{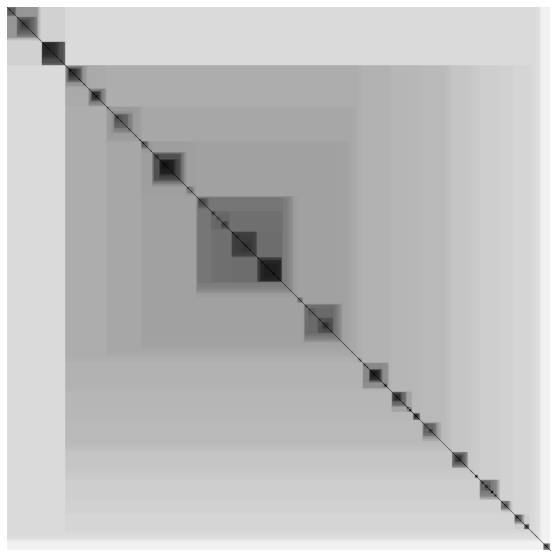}
        \caption{RN-30}
        \label{fig:RN-VAT30}
    \end{subfigure}

        \begin{subfigure}[b]{0.24\columnwidth}
        \centering
        \includegraphics[width=\columnwidth]{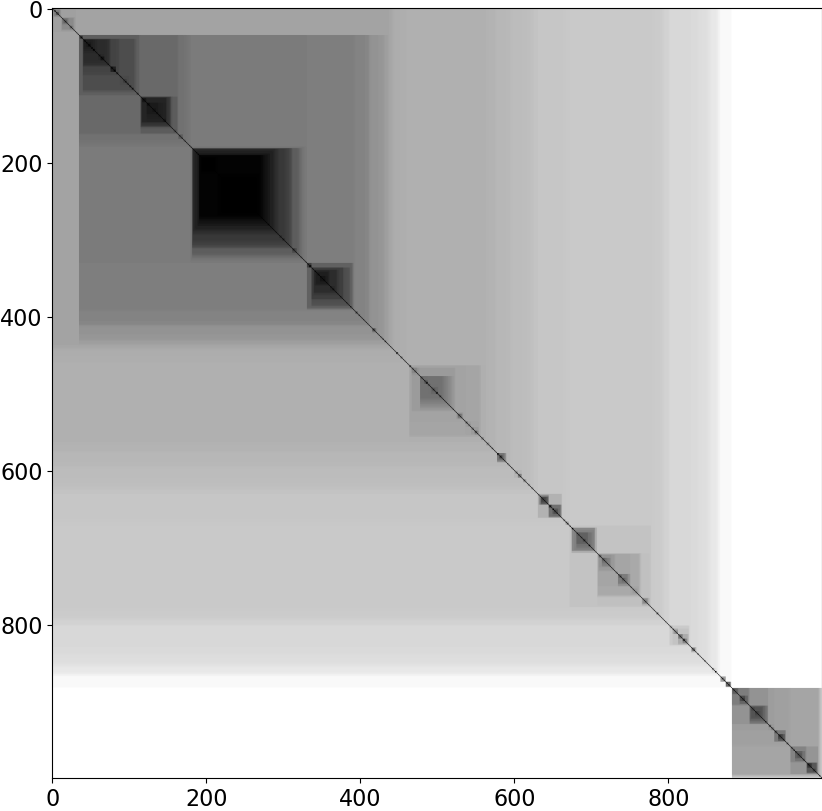}
        \caption{VGG-10}
        \label{fig:VGG-VAT10}
    \end{subfigure}
        \begin{subfigure}[b]{0.24\columnwidth}
        \centering
        \includegraphics[width=\columnwidth]{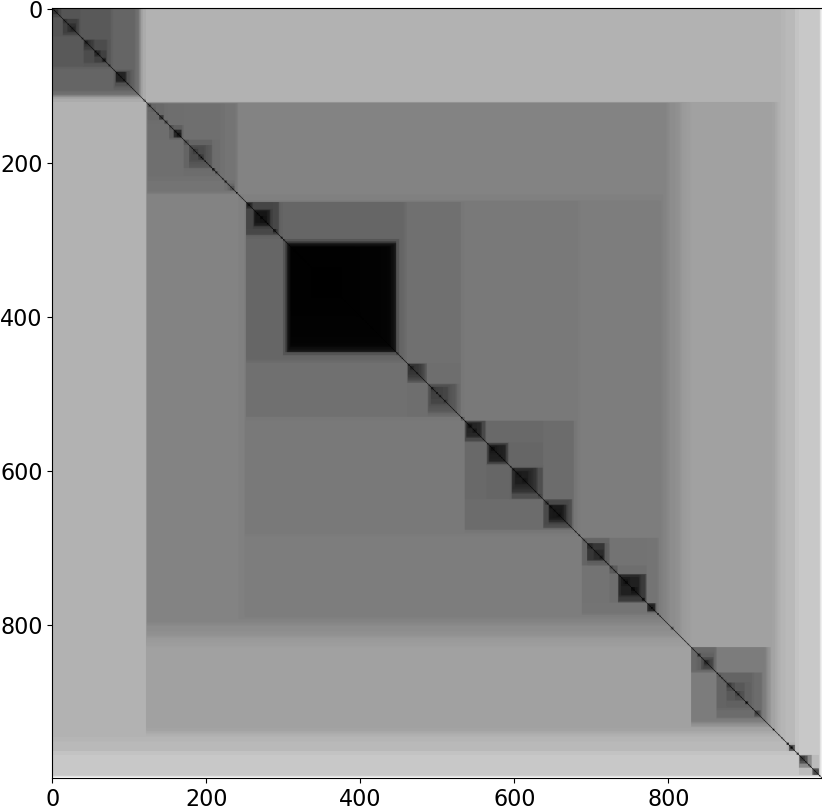}
        \caption{VGG-20}
        \label{fig:VGG-VAT20}
    \end{subfigure}
    \begin{subfigure}[b]{0.24\columnwidth}
        \centering
        \includegraphics[width=\columnwidth]{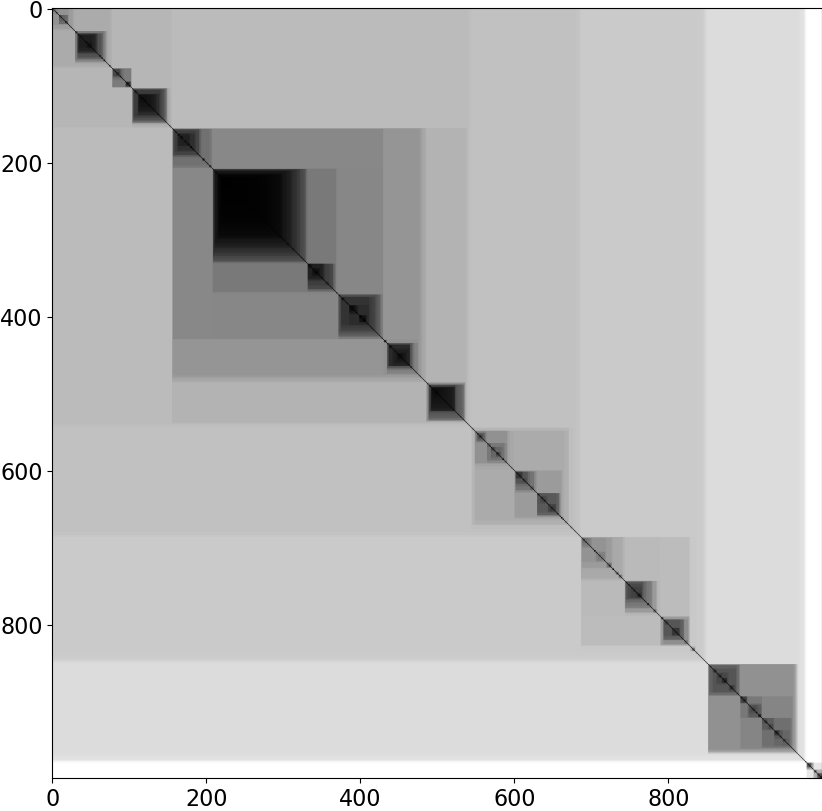}
        \caption{VGG-25}
        \label{fig:VGG-VAT25}
    \end{subfigure}
    \begin{subfigure}[b]{0.24\columnwidth}
        \centering
        \includegraphics[width=\columnwidth]{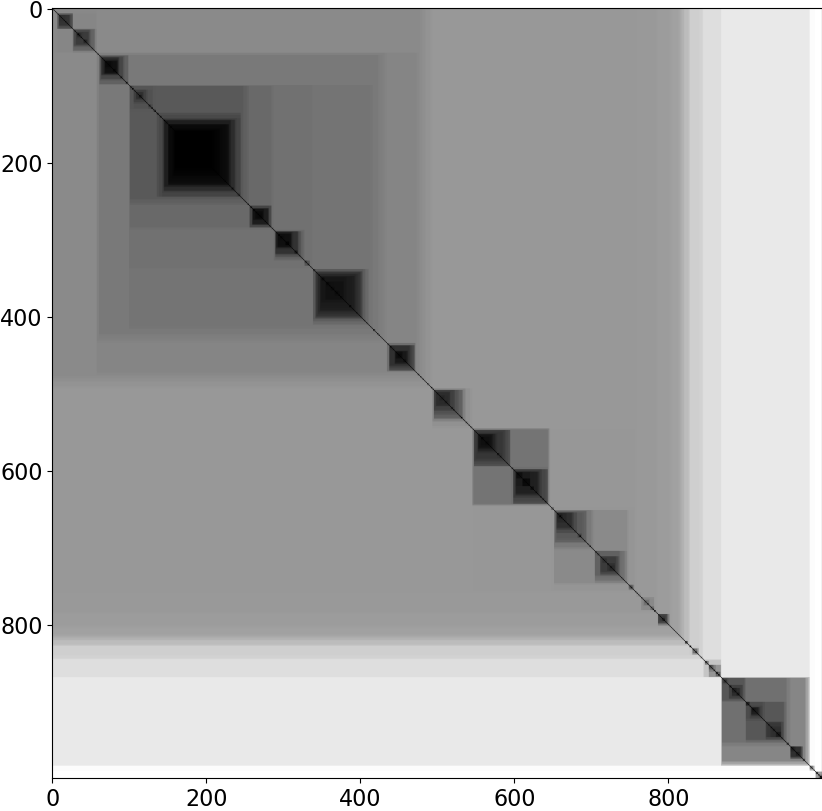}
        \caption{VGG-30}
        \label{fig:VGG-VAT30}
    \end{subfigure}
    
\caption{VAT plots for DL-based representations.}
\label{fig:DLbasedVAT}
\end{figure}

Figure~\ref{fig:DLbasedCumSum} shows EVR of the features learnt by the Representation learning block using ResNet18 (RN) and VGG11 (VGG) DL-based models with different number of clusters. A smaller number of clusters yields higher EVR in the lower components, while there is no significant difference in the cumulative sum of EVR after the 20$^{th}$ component across different models. All models encode features with more than 95\% of EVR within 27 components which is 0.7\% of the 3770 components required by the baseline PCA. 

The VAT plots for the proposed RN-based model are shown in Figure~\ref{fig:DLbasedVAT}b--e and for the VGG-based model in Figure~\ref{fig:DLbasedVAT}f--i. The plots correspond to 4 different automatic representation learning models, configured for 10, 20, 25 and 30 clusters. Both DL models achieve very similar EVR. Experimentally it was observed that models trained with lower number of clusters significantly worsen the clustering tendency according to VAT plots, so they were not considered in the subsequent analysis.
For a smaller number of clusters in Figure~\ref{fig:RN-VAT10}, the learnt representations with RN-based model contain less prominent dark squares compared to the models with larger number of clusters in Figures~\ref{fig:DLbasedVAT}b--c. Similar observation holds for VGG-based models in Figures~\ref{fig:DLbasedVAT}e--h. For both DL models, the 25 clusters models show the most distinguished separation of the feature space.
This analysis indicates that the proposed architecture is able to learn representations that can yield 11 to 25 well separated clusters. It is also able to learn 5 to 10 and 26 to 30 less clearly separated clusters while it is less suitable for small number of clusters such as 2 to 4. Overall, the automatically learnt representation is able to extract fine-grained clusters containing the shapes of the wireless transmission bursts.

Comparing to the baseline, the CNN-based model can learn to encode the relevant information for cluster development in only approximately 0.7\% of the components required by the PCA baseline, when the application requires higher numbers of well defined clusters, while also enabling superior cluster differentiation. For two clusters, the baseline model provides a better separation according to the VAT plots and the same feature dimensionality size.

\subsection{Cluster analysis}
\begin{figure}[htbp]
    \centering
    \includegraphics[width=\columnwidth]{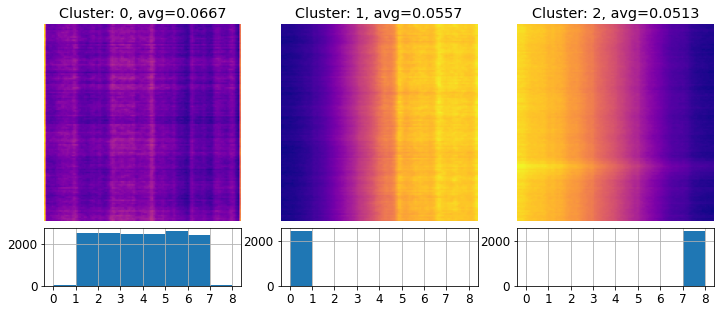}
    \caption{Distribution of samples from each cluster along the frequency band for the baseline approach.}
    \label{fig:SamplesBandDistribution}
\end{figure}

\begin{figure}[htbp]
    \centering
    \noindent\makebox[\columnwidth]{\includegraphics[width=\columnwidth]{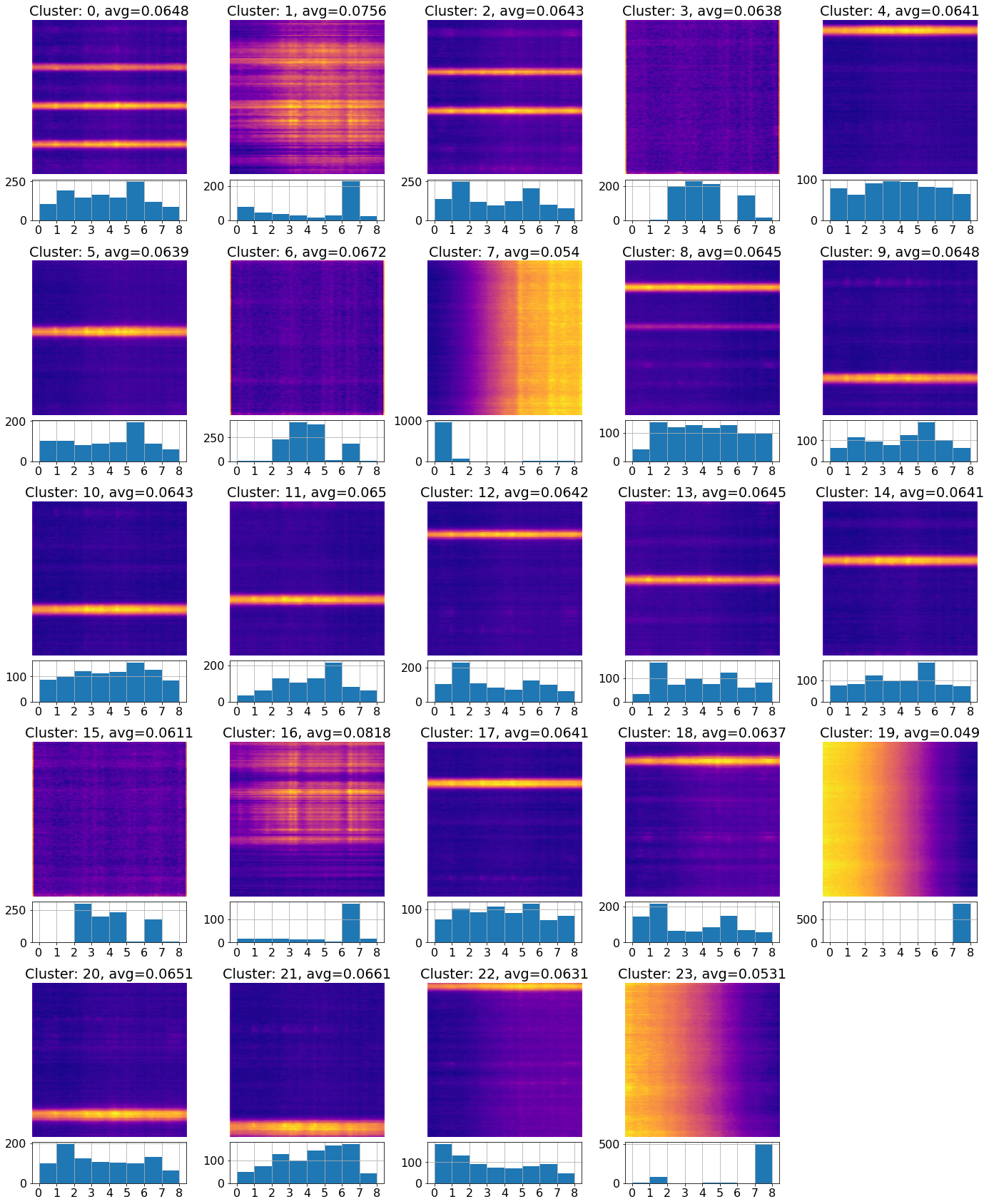}}
    
    \caption{Distribution of samples from each cluster along the frequency band for the clusters obtained by the RN-based approach.}
    \label{fig:SamplesBandDistributionAuto}
\end{figure}

Next we examined the best clusters developed with the baseline and the CNN-based approach using histograms of samples accompanied with the average cluster spectrograms in Figure~\ref{fig:SamplesBandDistribution} and Figure~\ref{fig:SamplesBandDistributionAuto}, respectively.

For the best baseline approach containing 3 clusters, Figure~\ref{fig:SamplesBandDistribution} shows that the cluster 0 contains mostly the samples from sub-bands between 1 and 7, the cluster 1 contains the samples from the left-most sub-band and the cluster 2 contains almost all of the samples from the right-most sub-band. Sub-bands refer to sections of frequency-wise segmentation as shown in Figure~\ref{fig:SampleData}. This observation is also aligned with the size of the dark squares in the VAT plot of the 2 PCA features in Figure \ref{fig:baselineVAT2}, which contains 3 clusters, one big corresponding to the cluster 0, and two almost equal smaller clusters corresponding to the clusters 1 and 2. Clearly, the baseline approach clusters the data based on the weaker signal on the left-most and right-most samples of the full bandwidth. The weaker signal seems to be a consequence of the nonuniform sensing capability of the sensor. Looking back to Figure \ref{fig:SamplesBandDistribution}, the left-most and the right-most samples have gradually vanishing brightness towards the edges. Plotting the average of the assigned spectrograms from each of the clusters supports this observation. 

Experimentally we identified that the 24-cluster automatic CNN-based model using ResNet18 DL architecture provides the best results. For this model, the average spectrograms and histograms in Figure~\ref{fig:SamplesBandDistributionAuto} show its effectiveness in learning general features related both to the transmission-specific content and the "background" of the spectrograms. The combined clusters 19 and 23 in Figure~\ref{fig:SamplesBandDistributionAuto} are the same as the cluster 2 from the baseline approach, occupying the right-most sub-band, while the cluster 7 corresponds to the cluster 1 of the baseline. According to Figure~\ref{fig:SamplesBandDistributionAuto}, the samples assigned to this cluster are again occupying the right-most sub-band. This means that the automatic model can also learn the features extracted with the baseline approach. Additionally, the automatic model learns features that are specific for the different patterns generated by the transmissions. The clusters 0, 2, 4, 5, 8, 9..14, 17, 18, 20..22, in Figure~\ref{fig:SamplesBandDistributionAuto} show horizontal line activities, which according to Figure~\ref{fig:SampleData} appear across the entire bandwidth. Their histograms show that samples assigned to these clusters are from all 8 sub-bands, and their distribution along the entire channel is roughly uniform. The clusters 3, 6 and 15 show the capability of the automatic model to distinguish the transmission-free spectrograms. This can be used to determine transmission-free sub-bands, which is another advantage against the baseline approach. Finally, the clusters 1 and 16 show groups of dot-like transmission bursts.

\begin{figure}[htbp]
    \centering
    \includegraphics[width=\columnwidth]{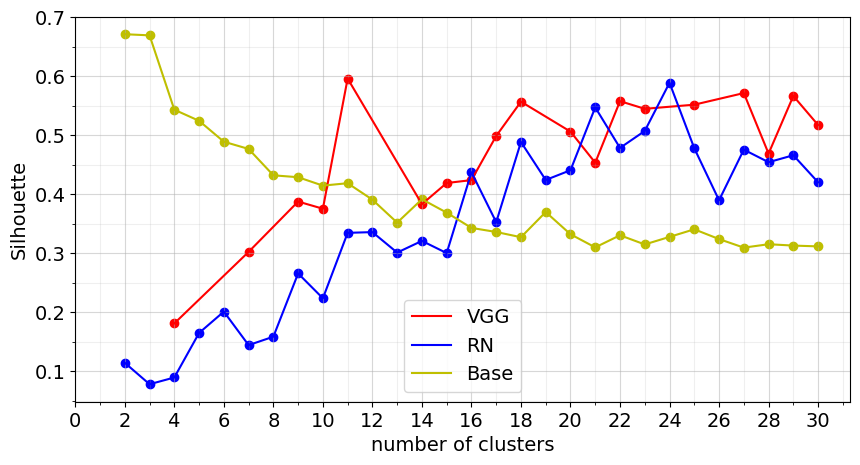}
    \caption{Silhouette scores}
    \label{fig:Silhouette}
\end{figure}

The Silhouette scores depicted in Figure~\ref{fig:Silhouette} for the baseline and the two CNN-based approaches confirm the observations based on the VAT plots. The baseline approach exhibits better performance at small number of clusters, for three clusters achieving the best score of 0.68. However, the baseline-provided feature space does not show transmission specific groups as samples are clustered based on the background noise. The automatic CNN-based models show comparable performance across the variation of clusters. This justifies the usage of the proposed lower complexity CNN, the ResNet18 instead the VGG11, preserving the performance and significantly reducing the complexity in terms of the number of required DL model parameters by roughly 11 times, according to Table~\ref{tab:complexity}. 

\begin{table}[h]
	\centering
	\footnotesize
        \caption{Complexity comparison.}
        \label{tab:complexity}
        \begin{tabular}{p{0.15\columnwidth}|p{0.15\columnwidth}p{0.15\columnwidth}p{0.15\columnwidth}}
		\toprule
            Algorithm & \cellcolor{gray!25} RN & VGG & Baseline \\
            \midrule
            Num. parameters & \cellcolor{gray!25} 11 M & 133 M & / \\
            \bottomrule
   	\end{tabular}	
\end{table}

\section{Conclusions}
\label{sec:conclusions}
In this paper, an automatic feature representation learning architecture based on CNN and PCA was explored and compared to a baseline model using only PCA for the task of clustering spectrograms from radio spectrum measurements. Our findings show that the baseline approach is useful when clustering based on general features of the data is required, with only a small number of clusters. On the other side, the automatic learning combining CNN and PCA, although more complex, provides much finer distinction between closely related groups of samples, based on their actual content, which are the transmissions bursts in our case. This shows that such architecture can be used for automatic representation learning and is suitable when large yet unlabeled spectrogram data is available. 

\section*{Acknowledgments}
This work was funded in part by the Slovenian Research Agency under the grant P2-0016 and in part by the European Union’s Horizon Europe Framework Programme under the grant agreement No 101096456 (NANCY).
The project is supported by the Smart Networks and Services Joint Undertaking and its members.

\bibliographystyle{IEEEtran}
\bibliography{main}

\end{document}